\documentclass[a4paper,twocolumn,
english,aps,pre,floatfix,groupedaddress,showpacs,nofootinbib]{revtex4-1}
\usepackage[T1]{fontenc}
\usepackage[latin1]{inputenc}
\usepackage{amsmath}
\usepackage{babel}
\usepackage{graphics}
\usepackage{amssymb}
\usepackage{mathrsfs}
\usepackage{dcolumn}


\begin{document}
\title
 {Domain wall theory and non-stationarity in driven flow with exclusion}
\author {R. B. \surname{Stinchcombe}}
\email{Robin.Stinchcombe@physics.ox.ac.uk}
\affiliation{Rudolf Peierls Centre for Theoretical Physics, University of
Oxford, 1 Keble Road, Oxford OX1 3NP, United Kingdom}
\author {S. L. A. \surname{de Queiroz}}
\email{sldq@if.ufrj.br}
\affiliation{Instituto de F\'\i sica, Universidade Federal do
Rio de Janeiro, Caixa Postal 68528, 21941-972
Rio de Janeiro RJ, Brazil}

\date{\today}

\begin{abstract} 
We study  the dynamical evolution toward steady state of the stochastic non-equilibrium model
known as totally asymmetric simple exclusion process, in both uniform and non-uniform (staggered) one-dimensional systems with open boundaries. 
Domain-wall theory and numerical simulations are used and, where
pertinent, their results are compared to existing mean-field predictions  and exact solutions
where available. For uniform chains we find
that  the inclusion of fluctuations inherent to the domain-wall formulation plays a crucial role
in providing good agreement with simulations, which is severely lacking in the corresponding mean-field predictions. For alternating-bond chains the  domain-wall predictions for the features of the
phase diagram in the parameter space of injection and ejection rates turn out to be realized 
only in an incipient and quantitatively approximate way. Nevertheless, significant quantitative agreement can be found between several additional domain-wall theory predictions and numerics.
\end{abstract}
\maketitle
 
\section{Introduction} 
\label{intro} 
In this paper we consider the dynamic evolution of the totally asymmetric simple exclusion
process (TASEP) in both uniform and non-uniform one-dimensional (1D) systems. 
The TASEP, in its original version for uniform chains, already exhibits
many non-trivial properties including flow phase changes, because of its
collective character~\cite{derr98,sch00,mukamel,derr93,rbs01,be07,cmz11}, and is considered 
paradigmatic of non-equilibrium statistical mechanical models.
We make use
of the domain-wall (DW) approach~\cite{ksks98,ps99,ds00,sa02}, coupled with numerical simulations. 

Application of mean-field (MF) Mobius mapping~\cite{hex13,hex15} to generalizations of
the TASEP such as non-uniform chains and hexagonal-lattice systems
turns out to provide  less accurate steady state results than for the well-known uniform 
1D case. And
for dynamics~\cite{hex15}, some significant discrepancies between MF predictions and
numerics already arise for uniform 1D systems.
Such inadequacies motivate the formulation of a DW theory 
(which includes fluctuations altogether absent in MF) to provide further understanding of the physical processes underlying this model.

In Sec.~\ref{sec:theo} we review the DW theory for uniform 
chains, and develop a generalization which applies for 1D systems with
alternating bond rates.  In Sec.~\ref{sec:num} we give results of the numerically
calculated DW evolution predicted by theory, as well as those from direct simulations of the
stochastic TASEP process. In Section~\ref{sec:conc}, we summarize and discuss our results.

\section{Theory}
\label{sec:theo}

\subsection{Uniform chain}
\label{sec:theo_unif_chain}
 We here briefly review basic aspects of the TASEP,  and of
its DW representation, for the simplest case, uniform 
chains~\cite{derr98,sch00,mukamel,derr93,rbs01,be07,cmz11,ksks98,ps99,ds00,sa02}.

In the TASEP the particle number $n_\ell$ at lattice site $\ell$ can be $0$ or $1$.
Any  such "exclusion" configuration of particles  (having at
most one particle at any site)  can evolve by hopping of the particle at any
occupied site $\ell$  to the adjacent site $\ell+1$, provided it is empty.
The instantaneous current $J_{\ell\,\ell+1}$
across the bond from $\ell$ to $\ell +1$ depends also on the stochastic attempt rate,
or bond (transmissivity) rate, $p_\ell$, associated with it. In the uniform TASEP, $p_\ell=p$
for all ``internal bonds'' $\ell$ (i.e. other than the injection and ejection ones for the  open-chain
case, see below). Thus,
\begin{equation}
J_{\ell\,\ell+1}= \begin{cases}{n_\ell (1-n_{\ell+1})\quad {\rm with\ probability}\ p_\ell}\cr
{0\qquad\qquad\qquad {\rm with\ probability}\ 1-p_\ell\ .}
\end{cases}
\label{eq:jinst}
\end{equation}

The "open" chain with additional  processes (subject to the exclusion
constraint) of injection at rate $\alpha$ at the left boundary, and
ejection at rate $\beta$ at the right boundary is of particular
interest, because of its rich behavior, including boundary-driven phase
transitions and associated static and dynamic critical behavior. Its
properties, particularly densities, currents, and associated correlation
functions,  will be studied in what follows.

One such property is the density profile $\rho (\ell,t)$ given by the average of
occupations at site $\ell$ and time $t$,  over all possible realizations of 
the evolution up to time $t$. In general such quantities evolve in time 
towards an eventual steady-state form which, for the open chain, depends on the 
boundary conditions through $\alpha$, $\beta$, but not on  initial conditions.

The simplest approach, MF theory~\cite{mukamel,rbs01}, already distinguishes the
different phases, through current and density averages and especially
through the forms of the density profile. Remarkably MF theory gives
the phase boundaries in the ($\alpha, \beta$) plane exactly for the uniform
chain~\cite{mukamel,rbs01}.

A particular example of an MF steady state density profile is that
for $\alpha = \beta< 1/2$, corresponding to the coexistence line in the low
current phase. This profile, for large system size, provides a
'macroscopic' view of the system state, in which a narrow domain wall
separates a domain on the left side, with uniform site occupation (local
density) $\rho^-$  controlled by the injection rate  from
another on the right with uniform site occupation $\rho^+$:
\begin{equation}
\rho^-=\alpha\quad;\qquad\rho^+=1-\beta\ ;
\label{eq:rhomrhop}
\end{equation}
and similarly for the mean field currents in the two domains.

However these steady state currents do not balance at the domain wall, if it
is stationary. This and other examples at different  ($\alpha, \beta$)
indicates the need to allow for (stochastic) motion of the domain wall.
This  is the motivation for the DW theory, which can restore the
particle conservation and include fluctuations absent from mean field
theory~\cite{ksks98,ps99,ds00,sa02}.  

One postulates that the TASEP process can be represented by the stochastic hopping of the
domain wall. For simplicity we make all bond rates $p_\ell \equiv 1$, see Eq.~\ref{eq:jinst}.

To be consistent with the particle currents near the wall one has to allow for possibly asymmetric
hopping rates  $D^+$, $D^-$ given by:
\begin{eqnarray}
D^-=\frac{j^-}{\Delta \rho}=\frac{\alpha(1-\alpha)}{1-\alpha-\beta}\ ; \nonumber \\
D^+=\frac{j^+}{\Delta \rho}=\frac{\beta(1-\beta)}{1-\alpha-\beta}\ .
\label{eq:dmdp}
\end{eqnarray}
Here  $\Delta \rho \equiv \rho^+-\rho^-=1-\alpha-\beta$, and
the currents $j^+$, $j^-$ within each domain are assumed~\cite{ksks98}
 to take the MF (i.e., factorized) form
\begin{eqnarray}
j^-=\rho^-(1-\rho^-)=\alpha(1-\alpha)\ ; \nonumber \\
j^+=\rho^+(1-\rho^+)=\beta(1-\beta)\ .
\label{eq:jmjp}
\end{eqnarray}
Despite the simplicity of the approach it does include fluctuations absent from the MF
picture and in some cases vastly improves on the MF description, e.g., in giving
certain exact results for the uniform chain (see, e.g., Sec.~\ref{sec:num}).

For a chain with $N$ sites and $L \equiv N+1$ bonds  (including the injection and ejection ones),
the time evolution of the probability $P(\ell,t)$ of finding the domain wall at "bond $\ell$"
(meaning the bond joining sites $\ell$ and $\ell+1$) for time $t$ is given by:
\begin{eqnarray}
\frac{dP(\ell,t)}{dt}=D^+ P(\ell-1,t)+D^-P(\ell+1,t) -\nonumber
\\ -\left(D^++D^-\right) P(\ell,t)\ ,
\label{eq:diffusion}
\end{eqnarray}
for internal bonds $1 \leq \ell \leq N-1$.
At the boundaries one has:
\begin{equation}
\frac{P(0,t)}{dt}=D^-P(1,t)-D^+P(0,t)\ ;
\label{eq:bounds1}
\end{equation}
\begin{equation}
\frac{dP(L,t)}{dt}=D^+P(L-1,t)-D^-P(L,t)\ .\quad\
\label{eq:bounds2}
\end{equation}
The general solution to Eqs.~(\ref{eq:diffusion})--(\ref{eq:bounds2}) can be found 
by assuming a linear superposition of forms $u^\ell\,e^{R(u)t}$.
Direct substitution into Eq.~(\ref{eq:diffusion}) shows that the following 
relation holds:
\begin{equation}
R(u)=\left(D^--\frac{D^+}{u}\right)(u-1)\ .
\label{eq:rvsu}
\end{equation}
So the steady-state solution $P_s(\ell)$, i.e. having $u$ such that $R(u)=0$, is
\begin{equation}
P_s(\ell)=c_1\,\left(\frac{D^+}{D^-}\right)^\ell +c_2\  ,
\label{eq:psst}
\end{equation}
thus  (for $D^+/D^- \neq 1$), the steady state density profile involves the  exponential factor $e^{\lambda_s\ell}$ where
\begin{equation}
\lambda_s=\ln \left(\frac{D^+}{D^-}\right)\ ,
\label{eq:lambda_s}
\end{equation}
corresponding to the wall being spread over a distance $\sim 1/|\lambda_s|$ at one
side of the system.
The time-dependent part of the full solution is formed by grouping together the degenerate factorizable solutions with $u$ and ${\bar u}=D^+/(D^-u)$ [$\,$such that $R({\bar u})=R(u)\,$]
into forms:
\begin{equation}
f(u,t)=\left[ A \,u^\ell+B\left(\frac{D^+}{D^-}\frac{1}{u}\right)^\ell\,\right] e^{R(u)t}\ .
\label{eq:pgen}
\end{equation}
The boundary conditions given in Eqs.~(\ref{eq:bounds1}) and~(\ref{eq:bounds2}) determine
 the allowed (discretized) $u$'s  ($\equiv u_n$) and the ratio of the coefficients $A$, $B$. So,
\begin{equation}
P(\ell,t)=\sum_n\left(A_n\,u_n^\ell+B_n\,{\bar u_n}^\ell\right)\,e^{R(u_n)t} +P_s(\ell)\ ,
\label{eq:pgen2}
\end{equation}
where
\begin{equation}
u_n=e^{\lambda_d}\,e^{iq_n}\ ;\quad {\bar u_n}=u_n^\ast\ ,
\label{eq:un}
\end{equation}
and
\begin{equation}
\frac{B_n}{A_n}=-\frac{(e^{\lambda_d}-e^{iq_n})}{(e^{\lambda_d}-e^{-iq_n})}\ ,
\label{eq:bnan}
\end{equation}
with
\begin{equation}
\lambda_d=\ln\sqrt{\frac{D^+}{D^-}}=\frac{1}{2}\lambda_s\ ;\quad  q_n= \frac{n\pi}{L} \ ,
\label{eq:ld_qn}
\end{equation}
and
\begin{equation}
R(u_n)=D^++D^--2\left[D^+D^-\right]^{1/2}\cos \frac{n\pi}{L} \equiv R_n\ .
\label{eq:r1}
\end{equation}
Once the probability $P(\ell,t)$ has been obtained, the density profile is given from:
\begin{equation}
\rho(\ell+1,t)-\rho(\ell,t)= \Delta \rho\,P(\ell,t)\ .
\label{eq:deltarho}
\end{equation}.

\subsection{Staggered chain}
\label{sec:theo_st_chain}

\begin{figure}
{\centering \resizebox*{2.8in}{!}{\includegraphics*{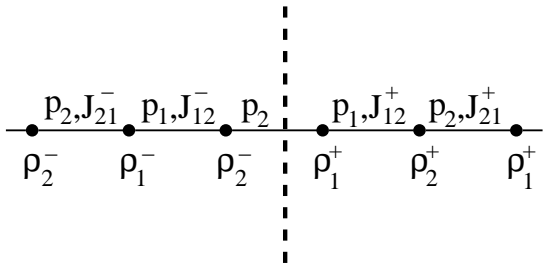}}}
\caption{
A domain wall (dashed vertical line) dividing a two-sublattice system  with
alternating hopping rates $p_1$, $p_2$, into a 'minus' domain (left)
 and a 'plus' one (see text for definitions of $\rho^{\pm}_{1,2}$ and 
$J^{\pm}_{12,21}$).}
\label{fig:dwalt_dpm}
\end{figure}

We next apply a DW approach to the TASEP with alternating bond rates. The geometry requires
a generalization of the usual macroscopic view, leading to new relationships of microscopic 
currents and densities to quantities such as diffusion rates. Of course macroscopic views
apply to each sublattice separately, but their interpenetration requires detailed consideration
of the particle current between sites on opposite sublattices. As usual in DW theory these
currents are those in the MF steady state, which are the same on all bonds (of either sublattice)
in a given domain. As in uniform chains with a domain wall, there remains the distinction 
between the uniform, MF steady state, densities $\rho^+$ and $\rho^-$ in domains on either
 side of the wall. But now these densities also differ between the two sublattices, which we distinguish by subscripts $1$ or $2$. 
The generalized hopping picture and the labeling on bonds of hopping rates
($p_1$,$ p_2$) and currents ($J_{12}^\pm$,$ J_{21}^\pm$), and of particle densities
($\rho_1^\pm$,$ \rho_2^\pm$) at sites, are shown in Fig.~\ref{fig:dwalt_dpm}.
The DW diffusion constants, resulting from particle conservation, for hopping to right or left ($+$ or
 $-$) from a given type of bond ($1$ or $2$)  are given by
\begin{equation}
 D^-_2\left(\rho^+_2-\rho^-_2\right)= J^-_{12} \equiv p_1\,\rho^-_1\left(1-\rho^-_2\right)\ ;
 \label{eq:d2m}
 \end{equation}
 \begin{equation}
 D^+_1\left(\rho^+_1-\rho^-_1\right)= J^+_{12} \equiv p_1\,\rho^+_1\left(1-\rho^+_2\right)\ ;
 \label{eq:d1p}
 \end{equation}
 \begin{equation}
 D^+_2\left(\rho^+_2-\rho^-_2\right)= J^+_{21} \equiv p_2\,\rho^+_2\left(1-\rho^+_1\right)\ ;
 \label{eq:d2p}
 \end{equation}
 \begin{equation}
 D^-_1\left(\rho^+_1-\rho^-_1\right)= J^-_{21} \equiv p_2\,\rho^-_2\left(1-\rho^-_1\right)\ .
\label{eq:d1m}
\end{equation}
Eq.~(\ref{eq:d2m}), for example, follows from the hopping picture and the labeling of 
bond rates and currents and site densities shown in Fig.~\ref{fig:dwalt_dpm}
because in the left hop of the wall shown, the bond left of the wall, with current $J_{12}^-$,
 carries  $\rho^+_2-\rho^-_2$ across to the right.
 
As for the uniform chain, the application of DW diffusion theory to the staggered chain needs the
 identification of incipient walls, typically from MF steady state density profiles, and their characterization. That involves finding hopping rates, using 
 Eqs.~(\ref{eq:d2m})--(\ref{eq:d1m}) with appropriate currents and densities. These can come
 from Mobius maps~\cite{hex13,hex15} involving the parameters $\alpha$, $\beta$, $p_1$, $p_2$.
 
 A variety of different forms arise, corresponding to the different regions of the MF steady state
 phase diagram. The MF phase boundaries, as well as coexistence and possible factorization lines, turn out to be only approximate for the staggered chain~\cite{hex15}, and they can be shifted
 by fluctuations. Their possible modification by DW diffusion is of particular interest and we first address that.
 
 For the staggered chain there is no known analogue of the operator algebra which holds
 for the uniform case, and from which the existence of factorizable states can be established,
 so here our use of the term "factorization" should be taken to cover the possibility of factorization
 through a state of uniform density. This issue will be discussed conclusively in 
 Sec.~\ref{sec:conc}.
 
 The DW diffusion steady state density profiles (on each sublattice) which determine such things
 are related to the steady state diffusion probability distribution $P_\ell$ for each sublattice,
 through a generalization of Eq.~(\ref{eq:psst}). As in the uniform
 chain, these $P_\ell$'s  typically include parts exponential in $\ell$.  The (coupled) steady state
 diffusion equations result in 
 \begin{equation}
 P_\ell = c_1\,\left(\frac{D^+_1D^+_2}{D^-_1D^-_2}\right)^{\ell/2} +c_2\  \qquad{\rm (steady\ state)}\  ,
 \label{eq:sstprof_st}
 \end{equation}
with different constants for the two sublattices.
 
 In the phase diagram the coexistence and factorization lines are special in having site-independent
 $P_\ell$'s. This can only occur if
 \begin{equation}
 \frac{D^+_1D^+_2}{D^-_1D^-_2}=1\ ,
 \label{eq:stgfac}
 \end{equation}
 analogous to having overall zero bias. 
 
 For converting the condition Eq.~(\ref{eq:stgfac}) to a relation between $\alpha$, $\beta$, $p_1$, 
$p_2$, one needs the MF steady state densities (uniform on each domain) for both  sublattices,
 namely $\rho^-_1$, $\rho^-_2$, $\rho^+_1$, $\rho^+_2$. With the injection and
 ejection sites both on sublattice $1$,
 \begin{equation}
 \rho^-_1=\frac{a p_2}{p_1+a(p_2-p_1)}\ ,\ \ 1-\rho^+_1=\frac{bp_1}{p_2+b(p_1-p_2)}\ ;
 \label{eq:abdef1}
 \end{equation}
 \begin{equation}
 \rho^-_2=a\  ,\qquad\quad\qquad\qquad1-\rho^+_2=b\   ,\qquad
 \label{eq:abdef2}
 \end{equation}
 where $a=\alpha/p_2$, $b=\beta/p_1$.
 
 With $A \equiv p_1+a(p_2-p_1)$, $B \equiv p_2+b(p_1-p_2)$, $C \equiv 1-a-b$, we find
 for the diffusion rates:
 \begin{eqnarray}
 D^+_1=b(1-b)\,\frac{A}{C}\; \nonumber \\ 
 D^+_2=p_1 p_2\,\frac{b(1-b)}{BC}\ ;\nonumber\\
 D^-_2=p_1 p_2\,\frac{a(1-a)}{AC}\ ;\nonumber\\
 D^-_1=a(1-a)\,\frac{B}{C}\ .
 \label{eq:d12pmf}
 \end{eqnarray}
 The sublattice density differences are
 \begin{equation}
 \rho^+_2 -\rho^-_2=C\ ;\qquad  \rho^+_1 -\rho^-_1= p_1 p_2\,\frac{C}{AB}\ ,
 \label{eq:drho1drho2}
 \end{equation}
and we have
 \begin{equation}
 \frac{D^+_1D^+_2}{D^-_1D^-_2}=\left[\frac{b(1-b)A}{a(1-a)B}\right]^2\ .
 \label{eq:stgfac2}
 \end{equation}
This last result makes the condition Eq.~(\ref{eq:stgfac}) for the coexistence and factorization lines, in DW theory, become
\begin{equation}
b(1-b)\left[p_1+a(p_2-p_1)\right]=\pm a(1-a)\left[p_2+b(p_1-p_2)\right]\ ,
\label{eq:cl_fac_cond}
\end{equation}
giving, respectively:
\begin{equation}
p_1\,\left(\frac{1-a}{a}\right)=p_2\,\left(\frac{1-b}{b}\right)\quad{\rm (coexistence)}\ ; \\
\label{eq:cl_dw_cond}
\end{equation}
\begin{equation}
a+b=1\qquad {\rm (factorization)}\ .
\label{eq:fac_dw_cond}
\end{equation}
It turns out that these determining equations are the same as in MF theory [$\,$where they come 
from the steady state equivalence of bond currents, with the uniform density profiles given by Eqs.~(\ref{eq:abdef1}),~(\ref{eq:abdef2})$\,$]. 

The rather general relations just given for diffusion rates and associated quantities can
need reinterpretation, e.g., to avoid sign errors in $D$'s, for certain regions of the phase diagram.

The intersection of the factorization and coexistence lines is the critical point
\begin{equation}
(a_c,b_c)=\left(\frac{\sqrt{p_1}}{\sqrt{p_1}+\sqrt{p_2}},\frac{\sqrt{p_2}}{\sqrt{p_1}+\sqrt{p_2}}\right)
\ ,
\label{eq:acbc}
\end{equation}
predicted by both MF and DW theory.

We next turn to dynamical behavior within DW theory, which needs use of the full coupled
discrete diffusion equations for the domain wall. Omitting the time dependence for clarity,
and recalling that odd-  and even-numbered lattice sites correspond respectively to sublattices
$1$ and $2$, these are:
\begin{equation}
\frac{dP_{2\ell}}{dt}=P_{2\ell-1}D^+_2+P_{2\ell+1}D^-_1-(D^+_2+D^-_1)P_{2\ell}\ ;
\label{eq:difcoup1}
\end{equation}
\begin{equation}
\frac{dP_{2\ell+1}}{dt}=P_{2\ell}D^+_1+P_{2\ell+2}D^-_2-(D^+_1+D^-_2)P_{2\ell+1}\  .
\label{eq:difcoup2}
\end{equation}
The solution of Eqs.~(\ref{eq:difcoup1})--(\ref{eq:difcoup2}) involves the following
two (sublattice) superpositions of factorizable components
\begin{equation}
P_{2\ell}=\sum_\zeta A_\zeta\,e^{2\ell\zeta}\,e^{-tR(\zeta)}\ ;
\label{eq:coupsol1}
\end{equation}
\begin{equation}
P_{2\ell+1}=\sum_\zeta B_\zeta\,e^{(2\ell+1)\zeta}\,e^{-tR(\zeta)}\ .
\label{eq:coupsol2}
\end{equation}
The resulting equations for $R(\zeta)$ and $B_\zeta/A_\zeta$ involve the matrix
\begin{equation}
{\cal M}=\begin{pmatrix}
{a(0)} & -a(\zeta)\cr  {-b(\zeta)} & {b(0)}\end{pmatrix}\ {\rm with}\
\begin{cases}{a(\zeta)=D^+_2 e^{-\zeta}+D^-_1 e^\zeta}\cr
{b(\zeta)=D^+_1 e^{-\zeta}+D^-_2 e^\zeta\ .}\end{cases}
\label{eq:matrix}
\end{equation}
The eigenvalues and eigenvectors of $\cal M$ provide a two-branch spectrum for $R(\zeta)$,
and corresponding values of the ratio  $B_\zeta/A_\zeta$. 

A few general remarks can be made here. One is that specifying $R(\zeta)=0$ requires that the
determinant of $\cal M$ should vanish, which is satisfied if $\zeta={\bar\zeta}$ where
\begin{equation}
e^{2{\bar\zeta}}=\frac{D^+_1 D^+_2}{D^-_1 D^-_2}\ .
\label{eq:zetabar}
\end{equation} 
This is the "complex wave vector" corresponding to the exponential profiles in steady state,
see Eq.~(\ref{eq:sstprof_st}).

Another remark concerns boundary conditions. These require that the differences of the
profiles from their steady state values have to vanish at the boundaries;
and they determine the allowed $\zeta$'s.

As in the uniform chain [$\,$see Eq.~(\ref{eq:pgen})$\,$] the boundary requirements can be satisfied
by grouping degenerate factorizable solutions, having $\zeta$'s with the same $R(\zeta)$. The
eigenvalue equation for $R(\zeta)$  is
\begin{equation}
R^2-\Sigma\,R+G(\zeta)=0\ , 
\label{eq:rquad}
\end{equation}
where
\begin{equation}
\Sigma=D^+_1+D^+_2+D^-_1+D^-_2 
\label{eq:sigdef}
\end{equation}
and, with
\begin{equation}
\Gamma=\sqrt{D^+_1D^+_2D^-_1D^-_2}\  ,
\label{eq:gammadef}
\end{equation}
\begin{equation}
G(\zeta)=2\Gamma\,\left(\cosh {\bar \zeta}- \cosh(2\zeta -{\bar \zeta})\right)\ .
\label{eq:gdef}
\end{equation}
So, degenerate $\zeta$'s  all have the same $G$, and a particular such group is easily seen to be
$z$, ${\bar \zeta}-z$, $z-\pi\,i$, ${\bar \zeta}+\pi\,i-z$, e.g. with $z$ real.

The generalization obtained by adding $\pm i\,q$ to each of these provides a group all with the same
${\rm Re}\ R$'s and equal or opposite ${\rm Im}\ R$'s (proportional to ballistic velocities).
Superpositions involving such a group provide the time-dependent parts of solutions which can
satisfy the boundary conditions. The remaining requirements are
\begin{equation}
z=\frac{\bar \zeta}{2}\ ;\qquad q=q_n=\frac{n\pi}{L}\ ,
\label{eq:zqn} \end{equation}
together with conditions relating the coefficients $A_\zeta$, $B_\zeta$ for all the $\zeta$'s of the
group. Initial conditions complete the determination of the coefficients.

The above procedure for dealing with the boundary conditions for the staggered chain is much
more complicated than that in Sec.~\ref{sec:theo_unif_chain}, but the result
Eq.~(\ref{eq:zqn}) is of the same form as Eq.~(\ref{eq:ld_qn}).

So the rate $R$ is provided by inserting $G(\frac{\bar \zeta}{2}\pm i\,q)$ into
 Eq.~(\ref{eq:rquad}).
Using Eqs.~(\ref{eq:zetabar})--(\ref{eq:gdef}) that gives
\begin{eqnarray}
G(\frac{\bar \zeta}{2}\pm i\,q)=\left(\sqrt{D_1^+D_2^+}-\sqrt{D_1^-D_2^-}\right)^2 +
\nonumber \\
+ 4\sqrt{D_1^+D_2^+D_1^-D_2^-}\sin^2q_n\ .
\label{eq:gfinal}
\end{eqnarray}
This provides the two-branch spectrum already referred to. In general $\bar \zeta$ is
nonzero and the spectrum has a gap, which is typically small.

Indeed the special case $D_1^\pm=D_2^\pm \equiv D^\pm$ has
\begin{equation}
G(\frac{\bar \zeta}{2}\pm i\,q)=\left(D^+-D^-\right)^2 +
4 D^+D^-\sin^2q_n\ ,
\label{eq:gfinal_u}
\end{equation}
which leads to the one-branch spectrum given in Eq.~(\ref{eq:r1}) as expected, since in this case 
the relationship of the $D$'s removes their sublattice distinction and so corresponds to the
uniform chain.

It can be seen from Eq.~(\ref{eq:zetabar}) that if $D_1^+D_2^+=D_1^-D_2^-$, corresponding to the
unbiased case [$\,$coexistence and factorization lines, see Eq.~(\ref{eq:stgfac})$\,$], $\bar \zeta$
becomes zero and $G=4D_1^+D_2^+ \sin^2 q_n$.
Then, at small $q_n$, $G$ becomes small and consequently
\begin{equation}
R \sim \frac{G}{\Sigma}\  ;\quad R \sim \Sigma -\frac{G}{\Sigma}\  ,
\label{eq:rzeta}
\end{equation}
for acoustic and optical branches respectively. The acoustic branch is gapless in this case,
which is analogous to the unbiased gapless case from $D^+=D^-$ in the uniform chain.

The acoustic branch provides the small-$q$ modes which dominate the late-time dynamical behavior.
The higher-$q$ modes of that branch and the modes of the other branch decay rapidly
 as in the "fast equalization" of sublattices previously studied in MF dynamics~\cite{hex15}.

In gapless cases or typical cases with a small gap the late-time modes have small $q$.
So, for these
 the ratio $B_\zeta/A_\zeta$ will be close to $a(0)/a({\bar\zeta}/2)$. This, being independent of
 $q_n$, makes $ (P_{2\ell+1}+P_{2\ell-1})/2P_{2\ell}$ independent of $\ell$.  This implies that, according to DW diffusion theory, at late times the
 DW distribution functions on the two sublattices are proportional.
 
 Now the densities on the sublattices can be found from the distribution functions using the following
 straightforward generalization of Eq.~(\ref{eq:deltarho}) for the uniform case:
 \begin{equation}
 \rho_{2\ell+1}(t)-\rho_{2\ell}(t)=(\rho^+_1-\rho^-_2)\,P_{2\ell}(t)\ ;
 \label{eq:diff1}
 \end{equation}
 \begin{equation}
 \rho_{2\ell}(t)-\rho_{2\ell-1}(t)=(\rho^+_2-\rho^-_1)\,P_{2\ell-1}(t)\ .
 \label{eq:diff2}
\end{equation} 
With these,  noting that in general $\rho^+_1-\rho^-_2 \neq \rho^+_2-\rho^-_1$,
the result above for the probability distributions becomes the statement that the late-time 
 difference of density profiles is very nearly constant in $\ell$.

\section{Numerics}
\label{sec:num} 

\subsection{Introduction}
\label{sec:num_intro}

For a chain with $N$ sites and $L=N+1$ bonds (including the injection and ejection 
ones), an elementary time step consists of $L$ sequential bond update attempts,
each of these according to the following rules: (1) select a bond at random, say, 
bond $ij$, connecting sites $i$ and $j$;
(2) if the chosen bond has an occupied site to its left and an  empty site to its 
right, then (3) move the particle across it with probability (bond rate)  $p_{ij}$.
If the injection or ejection bond is chosen, step (2) is suitably modified to
account for the particle reservoir (the corresponding bond rate being, respectively,  
$\alpha$ or $\beta$).

Thus, in the course of one time step, some bonds may be selected
more than once for  examination and some may not be examined at all.
This constitutes the {\it random-sequential update} procedure described
in Ref.~\onlinecite{rsss98}, which is the realization
of the usual master equation in continuous time~\cite{rsss98}.
For uniform chains the
exact steady-state profiles given by the operator algebra
described in Ref.~\onlinecite{derr93}, which are an important baseline in
our numerical work, correspond to
random-sequential update as recalled in  Ref.~\onlinecite{rsss98}.

For specified initial conditions, we generally took ensemble
averages of local densities and/or currents over $N_{\rm sam}=10^5$--$10^6$
independent realizations of stochastic update up to a suitable time
$t_{\rm max}$, for each of those collecting system-wide samples at
selected times.

Estimation of uncertainties involves running $N_{\rm set}$
independent sets of $N_{\rm sam}$ samples each; from the spread among the averaged 
quantities for the distinct sets, one then estimates the root-mean-square (RMS) 
deviation of each relevant quantity. As is well known~\cite{dqrbs96}, such RMS 
deviations are essentially independent of $N_{\rm set}$
as long as $N_{\rm set}$ is not too small, and vary as $N_{\rm sam}^{-1/2}$.
We generally took $N_{\rm set}=10$. Such stochastic fluctuations are the source
of the error bars displayed in 
Figs.~\ref{fig:dwa3b4ns},~\ref{fig:sstp_acbc_st},~\ref{fig:sstp_abcl_st}, 
and~\ref{fig:sstp_acbcb_st} below. 

\subsection{Uniform chain}
\label{sec:num_chain}

We started by testing the predictions of DW theory for selected steady-state properties of
uniform chains. In this case, the exact steady-state density profiles $\rho_s(\ell)$ are 
known~\cite{derr93} for any $(\alpha,\beta)$ and arbitrary number of sites $N$.

\begin{figure}
{\centering \resizebox*{2.8in}{!}{\includegraphics*{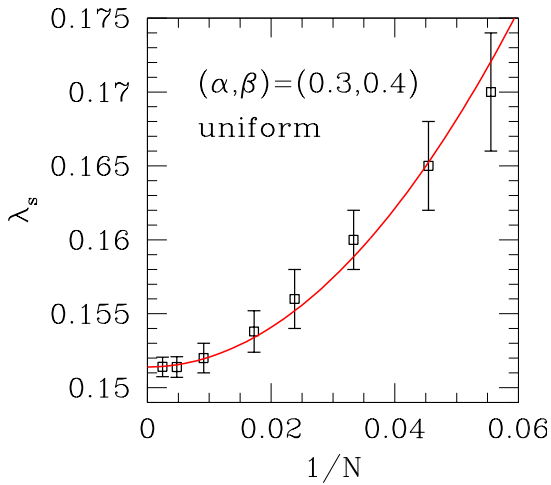}}}
\caption{
Adjusted values of $\lambda_s$ of Eq.~(\ref{eq:lambda_s}) for fits of
Eq.~(\ref{eq:rhostfit}) to exact steady-state profiles, for uniform chains of $N$ sites,
against $1/N$. Full line is a parabolic  spline through large-$N$ results  (see text).
}
\label{fig:ssta3b4}
\end{figure}

For $(\alpha,\beta)=(0.3,0.4)$, in which case   Eq.~(\ref{eq:dmdp}) gives $D^+=0.8$, $D^-=0.7$ 
we attempted to fit the exact profiles according to Eqs.~(\ref{eq:psst}), (\ref{eq:lambda_s}),
and~(\ref{eq:deltarho})
to the form
\begin{equation}
\rho_s(\ell)=a+b\,\exp(\lambda_s\,(\ell-\ell_0)) 
\label{eq:rhostfit}
\end{equation}
with $a$, $\lambda_s$, and $\ell_0$  as adjustable parameters; for $\alpha<\beta $ one  keeps
 $b=+1$ (fixed)  as is appropriate for  $\alpha+\beta <1$. 
Results for selected values of $N$  between $15$ and  $400$  are
displayed in Fig.~\ref{fig:ssta3b4},
where the uncertainties shown relate exclusively to the intrinsic features of  multiparametric
nonlinear regression.  
The quality of  fit improves for increasing $N$, as shown by the
shrinking standard deviations for $\lambda_s$; also, the central estimates tend to stabilize for $N \gtrsim 100$, suggesting  a parabolic form with no linear term in $N^{-1}$ to describe the asymptotic behavior for large $N$ 
[$\,$shown as a full (red) line in Fig.~\ref{fig:ssta3b4}$\,$]. This
gives $\lim_{N \to \infty} \lambda_s^{\rm fit}=0.151(1)$, to be compared with the prediction of Eq.~(\ref{eq:lambda_s}), $\lambda_s^{\rm DW}=0.13353 \dots$. 

For dynamics, we initially investigated the coexistence line (CL) between low- and high-density
phases, at $\alpha=\beta < 1/2$, where both $\lambda_s$ of Eq.~(\ref{eq:lambda_s})
and $\lambda_d$ of Eq.~(\ref{eq:ld_qn}) vanish.
In order to avoid crossover effects due to proximity to the
critical point at $(\alpha,\beta)=(1/2,1/2)$ we took $(\alpha,\beta)=(1/4,1/4)$. 
Keeping only the $n=1$ term in Eq.~(\ref{eq:pgen2}), the very-late time density difference 
profiles $\delta\rho(\ell,t) \equiv \rho(\ell,t)-\rho_s(\ell)$
behave, on the approach to steady-state, as
\begin{equation}
\delta\rho(\ell,t) = -\frac{2(1-2\alpha)}{\pi}\,\sin \frac{\pi \ell}{L}\,e^{-c(L)\,t}\ ,
\label{eq:rhocl}
\end{equation}
where 
the numerical prefactor comes from adjusting the $A_n$, $B_n$ of Eq.~(\ref{eq:pgen2})  
to an empty-lattice initial condition, and the inverse relaxation time is given, using Eqs.~(\ref{eq:dmdp}) 
and~(\ref{eq:r1}) for $L/\pi \gg 1$, by
\begin{equation}
c(L) = R_1(L)=\frac{\alpha(1-\alpha)}{1-2\alpha}\left(\frac{\pi}{L}\right)^2\ .
\label{eq:tau_cl}
\end{equation}
Note that Eq.~(\ref{eq:tau_cl}) coincides with the Bethe ansatz result of Ref.~\onlinecite{ess05}
[$\,$see their Eq.~(22)$\,]$.

\begin{figure}
{\centering \resizebox*{2.8in}{!}{\includegraphics*{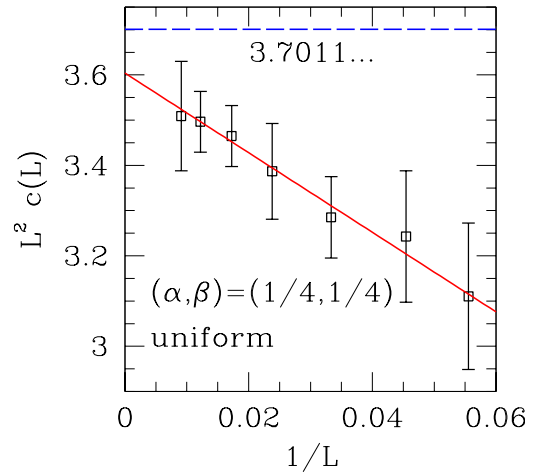}}}
\caption{
For $(\alpha,\beta)=(1/4,1/4)$, plot of $L^2\,c(L)$ against $1/L$, where the $c(L)$ are adjusted values of the exponential time decay of  $\{\delta\rho\}(L)$, see Eqs.~(\ref{eq:rhocl}) 
and~(\ref{eq:tau_cl}). The full (red) line is a linear fit to the data. The long-dashed line
marks the DW theory prediction.
}
\label{fig:cl_cl}
\end{figure}

For fixed $L$, we ran simulations starting from an empty lattice; then, for a set of suitable $t$
values we fitted numerically generated difference profiles to the sine dependence in 
Eq.~(\ref{eq:rhocl}), thus producing a sequence of effective time-dependent amplitudes,
which was in turn fitted to an exponential  time dependence to extract estimates of
the $c(L)$ of Eq.~(\ref{eq:tau_cl}). Finally, we examined the behavior of the $\{c(L)\}$
against $L$. Results are shown in Fig.~\ref{fig:cl_cl}.

The error bars shown in the Figure result from the cumulative effects of: (i) statistical fluctuations
in the local densities for each specified $\ell$ and $t$, coming from the stochastic sampling process;
 (ii) intrinsic  uncertainties following from adjusting  difference profiles for fixed $t$ to a single  sine dependence while fully neglecting higher-order terms in Eq.~(\ref{eq:pgen2}) [$\,$see
 Eq:~(\ref{eq:rhocl})$\,$]; and (iii) additional intrinsic uncertainties related to assuming the time dependence of the 
 effective amplitudes found in (ii) to follow a single exponential form over a relatively
 broad time interval. We have seen that  (ii) and (iii) are of much larger quantitative importance than (i). For instance, the $c(L)$ of Fig.~\ref{fig:cl_cl} have uncertainties varying between $2$ and $5\%$,
 while relative fluctuations in  the associated difference densities $\delta\rho(\ell,t)$ are of order  $1\%$ or less (provided that one analyses sites not very close to the system edges, where the
 $\delta\rho$ approach zero). Similar considerations apply to the respective sources of the error bars
 exhibited  in Figs.~\ref{fig:r1a3b4},~\ref{fig:nsp_abncl}, and~\ref{fig:nsp_abclp}
 below.

One sees that the numerical estimates of $c(L)$ become closer to the prediction of 
Eq.~(\ref{eq:tau_cl}) with increasing $L$.  The extrapolated value  is
$\lim_{L \to \infty} L^2\,c(L)= 3.6(1)$, to be compared to $\alpha(1-\alpha)\pi^2/(1-2\alpha)
=3.7011 \dots$ for $\alpha=1/4$.

Next we examined the time evolution of difference densities for $\alpha+\beta<1$, 
away from the CL.  
In this case DW theory gives the late-time difference densities as
\begin{equation}
\delta\rho(\ell,t) \propto \exp(\lambda_d\ell)\,\sin \frac{\pi \ell}{L}\,e^{-R_1(L)\,t}\ ,
\label{eq:rhoablt1}
\end{equation}
with $\lambda_d$, $R_1(L)$ as given respectively in Eqs.~(\ref{eq:ld_qn}) and~(\ref{eq:r1}).

At $(\alpha,\beta)=(0.3,0.4)$, using Eqs.~(\ref{eq:dmdp}) and ~(\ref{eq:r1}) for $L/\pi \gg1$ one gets
$R_1=0.003337\dots + 7. 3857\dots/L^2$.   Again, this agrees  with the Bethe ansatz result of 
Ref.~\onlinecite{ess05} [$\,$see their Eq.~(20)$\,]$.

We produced numerical estimates of $R_1$
by implementing a procedure similar to that described above for the CL. 
In contrast to that case, $\lambda_d$ is now an additional quantity to be considered.
 It is known~\cite{hex15}
that the predicted shapes of late-time profiles are very sensitive to the presence of an exponential term in their spatial dependence. Thus, in order to concentrate on the analysis of time decay rates
we took $\lambda_d$ as an adjustable parameter. However, the following remarks are in order.
We saw that  (i) for fixed system size $L$, the best-fitting values from numerics systematically decreased for increasing times $t$ in the non-stationary regime; and (ii) while, for assorted $L$ and $t$ one  generally found $0.07 \lesssim \lambda_d^{\rm fit} \lesssim
0.10$, an average over $L$ of long-time extrapolations of the behavior referred to in (i) 
gives $\langle\lambda_d^{\rm fit}\rangle=0.06(1)$. This is to be compared with the prediction $\lambda_d^{\rm DW}=0.06676 \dots$, and [$\,$using $\lambda_d=(1/2)\lambda_s\,]$ also to the final result for 
fits of steady-state profiles to Eq.~(\ref{eq:rhostfit}), namely $(1/2)\lambda_s^{\rm fit}=0.0755(5)$.

Our results for numerical estimates of $R_1$ are shown in Fig.~\ref{fig:r1a3b4}.   
\begin{figure}
{\centering \resizebox*{2.8in}{!}{\includegraphics*{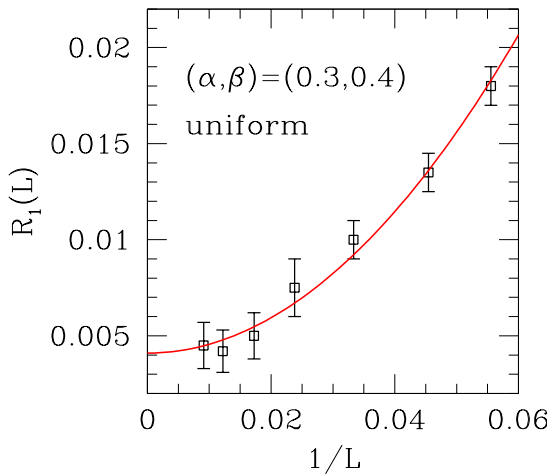}}}
\caption{
For $(\alpha,\beta)=(0.3,0.4)$, plot of $R_1(L)$ against $1/L$, where the $R_1(L)$ are adjusted values of the exponential time decay of  $\{\delta\rho\}(L)$, see Eqs.~(\ref{eq:r1})
and~(\ref{eq:rhoablt1}) . The full (red) line is a parabolic fit to the data (see text).
}
\label{fig:r1a3b4}
\end{figure}
The figure also shows a parabolic fit of the numerical data inspired in the large-$L$  limit
of Eq.~(\ref{eq:r1}). With $R_1(x)=R_1^{\,0}+R_1^{\,2}\,x^2$ one gets $R_1^{\,0}=0.0041(4)$,
 $R_1^{\,2}=4.6(2)$, the former value being only two error bars away from the DW prediction. 
 
Still for $(\alpha,\beta)=(0.3,0.4)$ we compared both the stationary and non-stationary behavior of density profiles, as given by DW theory, with corresponding results from, respectively,
the exact steady state solution and numerical simulations. To this end,
we solved the discrete-time version of Eq.~(\ref{eq:diffusion}),
\begin{eqnarray}
P(\ell,t+dt)=D^+dt\, P(\ell-1,t)+D^-dt\,P(\ell+1,t) + \nonumber
\\ +\left[1-(D^++D^-)dt\right] P(\ell,t)\ ,\qquad
\label{eq:diffdt}
\end{eqnarray}
with similar adaptations to Eqs.~(\ref{eq:bounds1}),~(\ref{eq:bounds2}).  Fixing $dt$ amounts
to a simple renormalization of the computational time scale with the proviso that the condition
$(D^++D_-)dt <1$ must be obeyed, to prevent negative probabilities cropping up upon iteration.
We used $dt=0.5$, which suffices for the present case. 

The density profiles can be evaluated at all times via~\cite{sa02}
\begin{equation}
\rho^{DW}(\ell,t)=\left(\sum_{k=0}^\ell P_k(t)\right)\,\rho^+ + \left(\sum_{k=\ell+1}^L P_k(t)\right)\,\rho^-
\ ,
\label{eq:prof_t}
\end{equation}
with $\rho^+$, $\rho^-$ from Eq.~(\ref{eq:rhomrhop}).

\begin{figure}
{\centering \resizebox*{2.8in}{!}{\includegraphics*{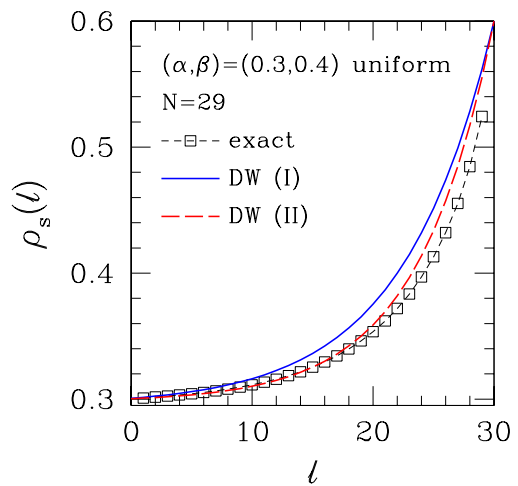}}}
\caption{
For $(\alpha,\beta)=(0.3,0.4)$, $N=29$, points are exact steady-state density profiles~\cite{derr93};
lines are results of long-time evolution of Eqs.~(\ref{eq:diffusion})--(\ref{eq:bounds2}). (I): $D^+$,
$D^-$ from Eq.~(\ref{eq:dmdp}). (II): $D^+/D^-=e^{\lambda_s}$, $\lambda_s=0.160$ (see text).
}
\label{fig:dwa3b4sst}
\end{figure}

Fig.~\ref{fig:dwa3b4sst},  for a system with $N=29$ sites,
shows the exact steady state profile~\cite{derr93} compared with two variants of 
the long-time limit of the evolution of Eq.~(\ref{eq:diffdt}) and the corresponding  versions of 
Eqs.~(\ref{eq:bounds1}),~(\ref{eq:bounds2}). In curve (I) we used $D^+$ and $D^-$ 
following Eq.~(\ref{eq:dmdp}), while in curve (II) we took $D^+/D^-=e^{\lambda_s}$,
with $\lambda_s=0.160$ being the central estimate from  the fit of the $N=29$ exact profile 
to Eq.~(\ref{eq:rhostfit}).

One sees that in both cases, although the general trends are captured by DW results, some
small but significant discrepancies remain especially close to the system's right end. One
expects such effects to become less relevant with increasing system size~\cite{sa02}.

\begin{figure}
{\centering \resizebox*{2.8in}{!}{\includegraphics*{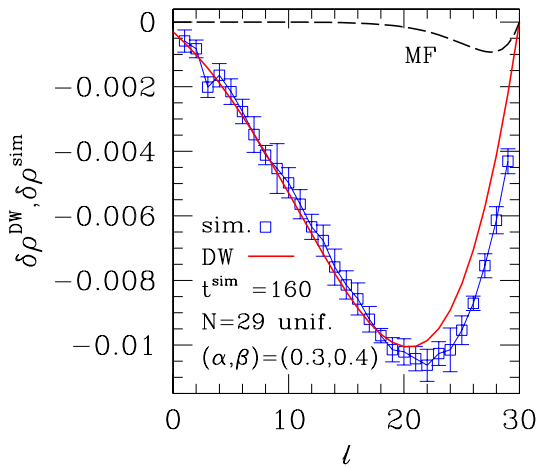}}}
\caption{
For $(\alpha,\beta)=(0.3,0.4)$, $N=29$, difference-density profiles. Points are simulation
results at $t^{\,\rm sim}=160$. Continuous (red) line is result of evolution of Eqs.~(\ref{eq:diffusion})--(\ref{eq:bounds2})
at $t^{\rm DW}=2t^{\,\rm sim}$. The long-dashed line is from MF theory of Ref.~\cite{hex15}.
See text for description of initial conditions.}
\label{fig:dwa3b4ns}
\end{figure}

We examined the approach to stationarity, by evaluating the difference densities
predicted by DW theory, i.e., $\delta\rho^{\rm DW}(\ell,t) \equiv \rho^{\rm DW}(\ell,t)- 
\rho_s^{\rm DW}(\ell)$. In Fig.~\ref{fig:dwa3b4ns} they are compared to those coming from simulations.
As mentioned previously, the latter use the exact steady state profiles as the baseline
to be subtracted from finite-time numerical results.

We  started the DW evolution with the domain wall at the right end of the system;
consistently with this, the numerical simulation was started with uniform average
density $\langle \rho \rangle=\rho^-=0.3$. With the elementary time step $dt=0.5$ for
the DW evolution, as mentioned, the correspondence between times scales is
 $t^{\rm DW}=2t^{\,\rm sim}$.  The features shown in Fig.~\ref{fig:dwa3b4ns} turn out to be typical
 of late-time profiles (say $100 \lesssim t^{\,\rm sim} \lesssim 250$), 
 namely the very good agreement between  DW and
 simulation results for $\ell \lesssim 20$, and the small but significant mismatch on the
 upturn for larger $\ell$, with the $\delta\rho^{\rm DW}$ profile approaching zero faster than
 that given by simulation.  The dashed line in Fig.~\ref{fig:dwa3b4ns} shows the 
 corresponding difference profile predicted by the MF theory of Ref.~\onlinecite{hex15}.
It is seen that there is everywhere a large discrepancy between the latter and simulation results.
For a similar (but simpler) case, namely $(\alpha,\beta)=(0.3,0.7)$, see Fig.~7 of
Ref.~\onlinecite{hex15}.

\subsection{Staggered chain}
\label{sec:num_st_chain}

We consider chains with alternating rates $p_1=1/2$, $p_2=1$ for all internal bonds (i.e.,
excluding the injection and ejection ones). The ratio $p_2/p_1=2$ is of special interest
since its mean-field Mobius mapping description coincides with that of a hexagonal-lattice
nanotube with uniform bond rates~\cite{hex13,hex15}. For consistency with the condition
expressed above Eqs.~(\ref{eq:abdef1}) and~(\ref{eq:abdef2}), the total number $N$ of sites
must be odd. 

\begin{figure}
{\centering \resizebox*{2.8in}{!}{\includegraphics*{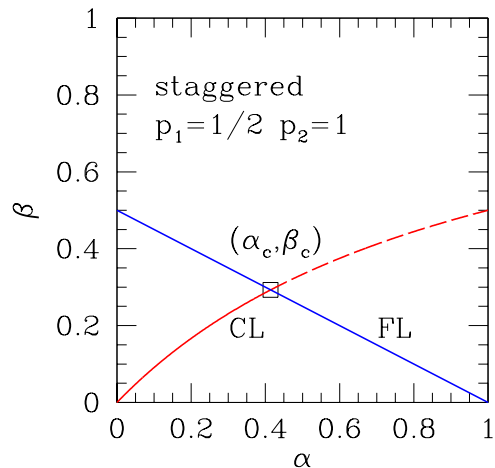}}}
\caption{
Phase diagram predicted by both MF and DW theory for staggered chain with bond rates $p_1=1/2$,
$p_2=1$; see Eqs.~(\ref{eq:clp1p2051})--(\ref{eq:cpp1p2051}) and text. CL denotes coexistence line; FL is factorization line. 
}
\label{fig:pd_st}
\end{figure}

Eqs.~(\ref{eq:cl_dw_cond}), (\ref{eq:fac_dw_cond}), and~(\ref{eq:acbc}) give, for $p_1=1/2$, $p_2=1$:
\begin{equation}
\beta=\frac{\alpha}{1+\alpha}\quad{\rm (coexistence)}\ ;
\label{eq:clp1p2051}
\end{equation}
\begin{equation}
\alpha+2\beta=1\quad{\rm (factorization)}\ ;
\label{eq:facp1p2051}
\end{equation}
\begin{equation}
(\alpha_c,\beta_c)=\left(\sqrt{2}-1,1-\frac{\sqrt{2}}{2}\right )\quad{\rm (critical\ point)}\ .
\label{eq:cpp1p2051}
\end{equation}
As explained in Sec.~\ref{sec:theo_st_chain}, the above are concurrent predictions from
MF and DW theory. Fig.~\ref{fig:pd_st} shows the overall features of the 
predicted phase diagram in  the $\alpha-\beta$ parameter space. In line with the uniform case,
one does not expect the continuation of the CL beyond $(\alpha_c,\beta_c)$ [$\,$long-dashed line in
Fig.~\ref{fig:pd_st}$\,$] to have a physical interpretation. 

An important feature of driven asymmetric flow on staggered chains~\cite{hex15} is
that no exact results for steady-state profiles or currents are known, e.g., from operator algebra, 
unlike the case of  their uniform-system counterparts~\cite{derr93}; thus guidance must come from 
numerically-generated data.
Nevertheless, some general properties which are known for uniform chains are expected to hold here as well, such as the existence of a low-current phase for suitably low
($\alpha$, $\beta$) and a high-current one for ($\alpha$, $\beta$) large enough.
For example the steady state current at $(\alpha,\beta)=(1/5,1/6)$, approximately halfway along the predicted CL, is $J \simeq 0.13$. To determine the maximal current  $J_{\rm max}$, 
we considered  the simpler case of staggered chains with periodic boundary conditions (rings), for which particle-hole symmetry arguments show that  $J_{\rm max}$  corresponds to a site-averaged
density $\langle \rho\ \rangle=1/2$. From numerical simulations for system sizes  with
$\langle \rho \rangle=1/2$, $N=20$, $30$, $40$
one gets $\lim_{N \to \infty} J_{\rm max}(N)=0.1628(1)$, to be compared with the
MF prediction $J_{\rm max}^{MF}=p_1p_2/(\sqrt{p_1}+\sqrt{p_2})=0.17157 \dots$. 

\begin{figure}
{\centering \resizebox*{2.8in}{!}{\includegraphics*{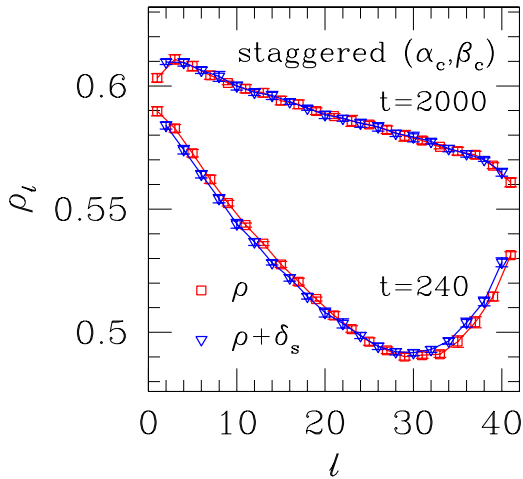}}}
\caption{
Sublattice density profiles for staggered chain with $N=41$ sites, at ($\alpha_c,\beta_c)$ of 
Eq.~(\ref{eq:cpp1p2051}), starting at $t=0$ with empty
lattice. The profile for $t=2000$ corresponds to  steady state regime to very good accuracy.
Densities for even-numbered sites (blue triangles) have been shifted upwards by $\delta_s=0.126$.
}
\label{fig:sstp_acbc_st}
\end{figure}

The verification of constant $\ell$-independent difference between sublattice 
steady state profiles, predicted in Eqs.~(\ref{eq:diff1})--(\ref{eq:diff2}), is illustrated in 
Fig.~\ref{fig:sstp_acbc_st}. Note that already for $t=240$ there is a good degree of
convergence towards a constant difference between sublattice profiles, although some
systematic and significant discrepancies still survive. 
 This  feature has been found to hold generally for various $(\alpha,\beta)$
spanning all expected phases. The numerical values of the difference $\delta_s$ (as defined in
Fig.~\ref{fig:sstp_acbc_st}) vary in the range $0.04 \lesssim \delta_s \lesssim 0.15$.

We first examined steady-state properties at points well within the predicted low-current
phases, i.e. for small $\alpha$, $\beta$ and suitably far from the predicted
CL $\beta=\alpha/(1+\alpha)$, $\alpha < \alpha_c=\sqrt{2}-1$.
We took (i) $(\alpha,\beta)=(0.1,0.22)$ and (ii)  $(\alpha,\beta)=(0.3,0.115)$.
In both cases DW theory predicts an exponential shape for the sublattice steady-state
densities, amenable to fitting via  Eq.~(\ref{eq:rhostfit}) above, with $b=+1$
for the former, and $b=-1$ for the latter (as they are located on opposite
sides of the CL). 

From Eqs.~(\ref{eq:sstprof_st}) and~(\ref{eq:d12pmf}) one gets, respectively,
$\lambda_s=0.65777 \dots$ for (i), and $-0.479014 \dots$ for (ii).
Fitting steady-state profiles from sublattice $1$ for chains with $N=41$ sites to   
Eq.~(\ref{eq:rhostfit}) gives, respectively, 
$\lambda_s^{\rm fit}=0.645(15)$ for (i), $-0.594(8)$ for (ii).  
Motivated by the uniform-chain case depicted in
Fig.~\ref{fig:ssta3b4}, in both cases we checked for a systematic  $N$-dependence
of $\lambda_s^{\rm fit}$. We took $N=29$ and $57$. 
For both $(\alpha,\beta)$ pairs the adjusted parameters stay within at most
two error bars from the corresponding $N=41$ values quoted above. 
So one can conclude that for the former case there is very good agreement
between theory and simulation, while in the latter a discrepancy of order $20\%$
is present.
 
\begin{figure}
{\centering \resizebox*{2.8in}{!}{\includegraphics*{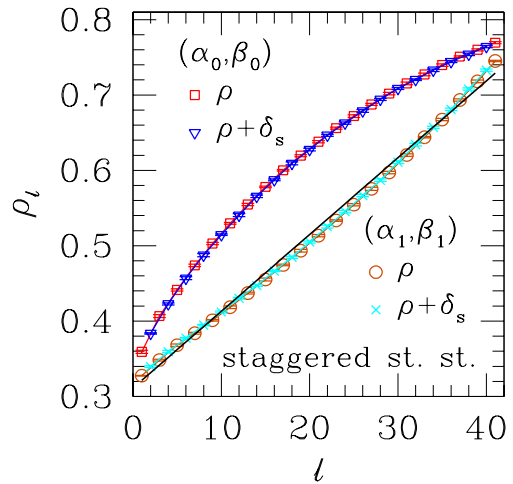}}}
\caption{
Steady state sublattice density profiles for staggered chain with $N=41$ sites, at
 ($\alpha_0,\beta_0)=(1/5,1/6)$  $[\,$on the predicted CL, see Eq.~(\ref{eq:clp1p2051})$\,$], 
 and $(\alpha_1,\beta_1)=(\alpha_0-2\varepsilon,\beta_0+\varepsilon)$, $\varepsilon=0.005$.
 Density values at points on sublattice 2 are shifted upwards by $\delta_{s0}=0.102$,
 $\delta_{s1}=0.1045$. Straight line is a fit to $(\alpha_1,\beta_1)$ data.
}
\label{fig:sstp_abcl_st}
\end{figure}

For uniform chains, the coexistence of low-and high- density phases on the CL can be directly
observed, see e.g. Fig.~8 of Ref.~\onlinecite{dqrbs08}; a secondary characteristic of the
CL is that the steady-state (ensemble-averaged) density profile is, to a very good approximation, 
linear~\cite{derr93} on it. We have probed the existence of the latter feature for staggered chains,
by scanning the $(\alpha,\beta)$ parameter space near the predicted CL. Fig.~\ref{fig:sstp_abcl_st}
shows steady state profiles for both $(\alpha_0,\beta_0)=(1/5,1/6)$
(on the predicted CL) and at  $(\alpha_1,\beta_1)=(\alpha_0-2\varepsilon,\beta_0+\varepsilon)$,
$\varepsilon=0.005$. It is seen that at the latter point one gets a rather good fit to a straight-line 
profile,  while there is pronounced curvature at the former. The straight line shown is a least-squares
fit to the  $(\alpha_1,\beta_1)$ data.

\begin{figure}
{\centering \resizebox*{2.8in}{!}{\includegraphics*{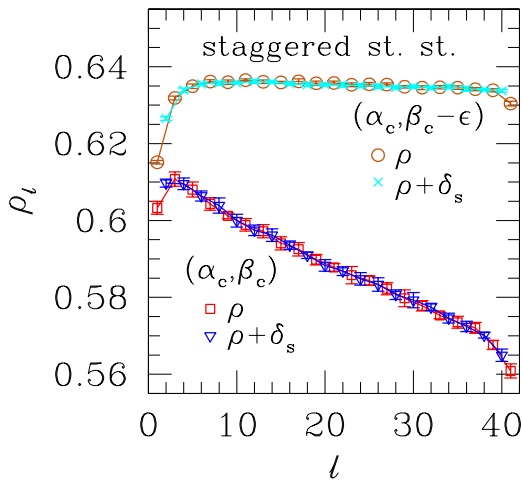}}}
\caption{
Steady state sublattice density profiles for staggered chain with $N=41$ sites, at the
predicted critical point $(\alpha_c,\beta_c)$  $[\,$see Eq.~(\ref{eq:cpp1p2051})$\,$], 
 and $(\alpha_c,\beta_c-\varepsilon)$, $\varepsilon=0.04$.
 Density values at points on sublattice 2 are shifted upwards by $\delta_{s}=0.126$,
 $\delta_{s}=0.123$ respectively for $(\alpha_c,\beta_c)$, $(\alpha_c,\beta_c-\varepsilon)$ . 
}
\label{fig:sstp_acbcb_st}
\end{figure}

Similarly, for the predicted factorization line on staggered chains, it has been shown by direct evaluation ~\cite{hex15} that  the corresponding correlation functions do not vanish there. For further discussion of this point, see Sec.~\ref{sec:conc} and Appendix~\ref{sec:app1}.
However, we have seen that a secondary feature, in this case the flatness of steady-state density profiles which holds exactly in the uniform case,
can be approximately found close to the predicted location of the critical point, as illustrated in
Fig.~\ref{fig:sstp_acbcb_st}. Leaving out the three leftmost sites, and the rightmost one,
on sublattice $1$  of the profile corresponding  to $(\alpha_c,\beta_c-\varepsilon)$, which are strongly influenced by the boundary conditions at the chain's ends, one has 
a gentle slope for the central section, amounting to a $0.3\%$ density variation in all. 
This is to be compared with the $6\%$  difference  found for the same section of the chain at   
$(\alpha_c,\beta_c)$.

For dynamics we used similar procedures to those of Sec.~\ref{sec:num_chain}, with
pertinent adaptations.
For chains with $N=17$, $21$, $29$, $41$, $57$, $81$, and $109$ sites [$\,L=N+1$ bonds$\,$] and 
late times  we evaluated the difference densities $\delta \rho(\ell,t)= \rho(\ell,t)-\rho_s(\ell)$.
In order to prevent lingering effects of the sublattice fast-equalization process from introducing
systematic distortions, we restricted ourselves to sites on sublattice $1$ (odd-numbered). 
 For each of a number (between $5$ and $10$) of suitable sites $\ell$ along the chain, and a set of suitably late times for each site, we fitted the
simulation data to a single exponential, thereby producing estimates of the rate $R_1=R_1(L)$ [$\,$see Eqs.~(\ref{eq:rquad})--(\ref{eq:gfinal}) with $n=1\,$]:
\begin{equation}
\delta\rho(\ell,t)=a(\ell)\,e^{-R_1t}\ .
\label{eq:expfit_st}
\end{equation}
In what folllows, the values used for the numerically-obtained  $R_1(L)$ are unweighted averages of the exponential-fit results over the several $\ell$'s used.

We first  investigated the approach to steady state in the neighborhood of the
predicted  CL, see Eq.~(\ref{eq:clp1p2051}), where one expects to find signatures of a gapless spectrum. Motivated by the shapes of steady-state density profiles shown in fig.~\ref{fig:sstp_abcl_st}, we took $(\alpha,\beta)=(\alpha_0-2\varepsilon,\beta_0+\varepsilon)$, with $(\alpha_0,\beta_0)=(1/5,1/6)$ on the predicted CL,  $\varepsilon=0.005$.
\begin{figure}
{\centering \resizebox*{2.8in}{!}{\includegraphics*{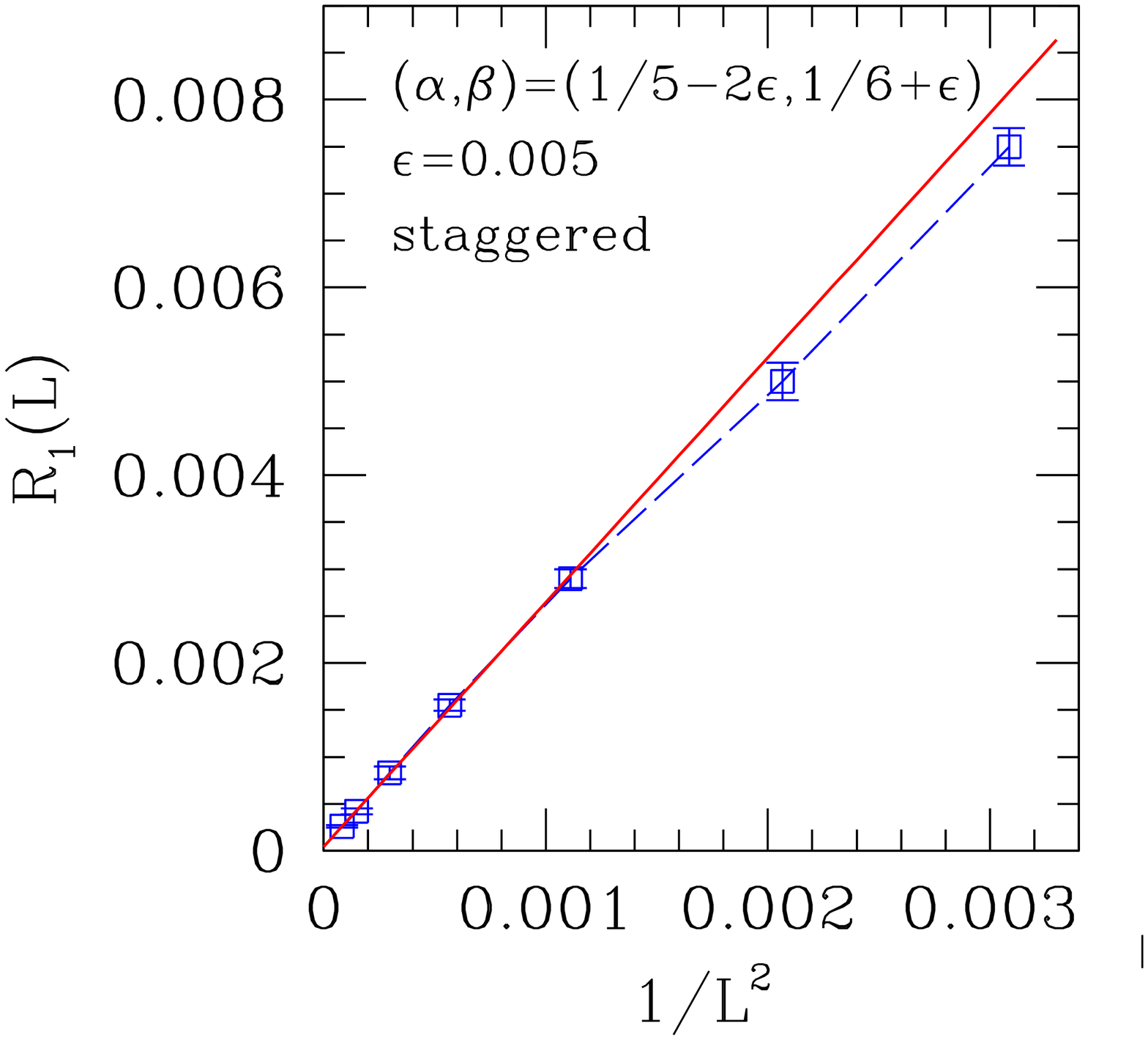}}}
\caption{
For $(\alpha,\beta)=(\alpha_0-2\varepsilon,\beta_0+\varepsilon)$, with $(\alpha_0,\beta_0)=(1/5,1/6)$ on the predicted CL of Eq.~(\ref{eq:clp1p2051}),  $\varepsilon=0.005$, plot of $R_1(L)$
against $1/L^2$
where points (blue squares) are adjusted values of the exponential time decay of
 $\delta\rho(\ell,t)$, see Eq.~(\ref{eq:expfit_st}). The full (red) line is a straight-line fit of numerical data for  $L \geq 30$ (see text).
 }
 \label{fig:nsp_abncl}
\end{figure}
The results for $R_1(L)$ are displayed in Fig.~\ref{fig:nsp_abncl}. The
full (red) line $R_1=aL^{-2}+b$ shown has adjusted parameters $a=2.60(3)$, $|b| < 4 \times 10^{-5}$.
One sees that for largeish $L \gtrsim 30$, essentially pure $1/L^2$ behavior has taken over, as attested by the smallness of $b$.
With the diffusion coefficients calculated from Eqs.~(\ref{eq:d12pmf}) and plugged into
Eqs.~(\ref{eq:rquad})--(\ref{eq:gfinal}) with $n=1\,$, one gets the predicted
gap to be $g_0=1.83 \times 10^{-4}$ at  $(\alpha_0-2\varepsilon,\beta_0+\varepsilon)$,
and $\lim_{L \to \infty}(g(L)-g_0)L^2=2.785 \dots$, the latter
to be compared with the adjusted slope $a$. If one uses instead the parameters at
$(\alpha_0,\beta_0)$, for which $g_0 \equiv 0$, the result is  $\lim_{L \to \infty}g(L)L^2=2.8196 \dots$.

\begin{figure}
{\centering \resizebox*{2.8in}{!}{\includegraphics*{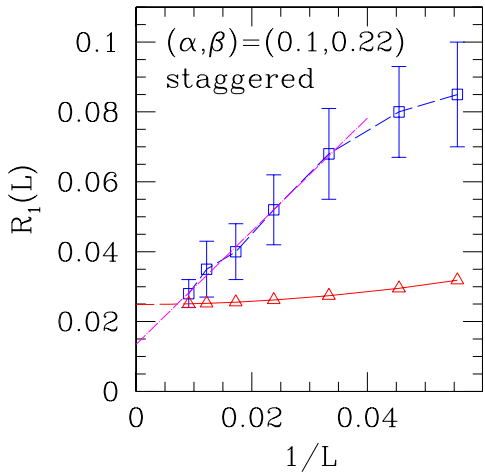}}}
\caption{
For $(\alpha,\beta)=(0.1,0.22)$, plot of $R_1(L)$ against $1/L$
where points (blue squares) are adjusted values of the exponential time decay of
 $\delta\rho(\ell,t)$, see Eq.~(\ref{eq:expfit_st}). The (red) triangles are the predictions of
 Eqs.~(\ref{eq:d12pmf}), together with Eqs.~(\ref{eq:rquad})--(\ref{eq:gfinal})~. The dot-dashed
 (magenta) line is a linear fit to large-$L$ numerical data (see text).
 }
 \label{fig:nsp_abclp}
\end{figure}

We also investigated $(\alpha,\beta)=(0.1,0.22)$, where
the agreement between theory and numerics for steady state profiles has proved to be very good
(see above). For these values of $(\alpha,\beta)$, Eqs.~(\ref{eq:d12pmf}) together with
Eqs.~(\ref{eq:rquad})--(\ref{eq:gfinal}) give $g_0=0.024907 \dots$, and 
 $\lim_{L \to \infty}(g(L)-g_0)L^2=2.2518 \dots$.  Our numerical results are shown in
 Fig.\ref{fig:nsp_abclp}.
An {\em ad hoc} linear fit of the five largest-$L$ data against $1/L$ (shown in 
Fig.~\ref{fig:nsp_abclp}) gives $g_0=0.013(2)$, albeit with
chi-squared per degree of freedom $(\chi^2_{\rm d.o.f})=0.03$ due to the rather broad error bars.
Assuming that, in qualitative agreement with theory, the asymptotic finite-size correction is in fact 
$A/L^2$ with $A>0$, the above extrapolation for $g_0$ can be seen as a loose lower bound for that
quantity.  

\section{Discussion and Conclusions} 
\label{sec:conc}

We initially discuss uniform chains. In general, the results of Sec.~\ref{sec:num_chain}
confirm that DW theory is a good approximation. It must be noted, however, that even for 
steady state some discrepancies remain: see the discussion of numerical data displayed in 
Figs.~\ref{fig:ssta3b4} and~\ref{fig:dwa3b4sst}. Of course this is because, although
the exact steady state profiles (away from factorization and coexistence lines) do behave to a large degree like the exponentials predicted by Eq.~(\ref{eq:psst}), they are not {\em identical} to
them. 

Regarding the approach to steady state, DW theory accurately predicts the existence and numerical value of the gap, at least in the low-current phase (and is in accordance 
with Bethe ansatz results~\cite{ess05} and simulations, including the main finite-size corrections); 
see Figs.~\ref{fig:cl_cl}  and~\ref{fig:r1a3b4}.
The good quantitative agreement between DW evolution and finite-time  simulations,
already illustrated in Ref.~\onlinecite{sa02}, is here highlighted  and given
further prominence by the stark
contrast of DW results with the sizable disagreement exhibited  by MF treatments against numerics, 
see Fig.~\ref{fig:dwa3b4ns} and Ref.~\onlinecite{hex15}.
Indeed, this strongly indicates that fluctuations (incorporated, albeit approximately, by DW theory, and ignored by MF treatments) are the crucial ingredient for the proper description of the
approach to steady state. 

For staggered chains, we recall that the features of the phase diagram predicted by DW theory coincide with those obtained from application of MF concepts. They are qualitatively similar to those 
established for uniform chains, exhibiting the special factorization- and coexistence 
lines along which steady-state density profiles are expected not to display
exponential behavior, see Eqs.~(\ref{eq:sstprof_st})--(\ref{eq:stgfac}),~(\ref{eq:cl_fac_cond}), and~(\ref{eq:cl_dw_cond}).
 
 Numerical evidence that factorization (as characterized by the vanishing of  the
 associated steady-state correlation functions) does not hold as predicted was already found in 
 Ref.~\onlinecite{hex15} for uniform-rate nanotubes. For such systems the MF Mobius
 mapping equations are identical to those for the staggered chain with $p_1/p_2=2$.
 Here we verified numerically that, similarly, correlations do not vanish for staggered
 chains, e.g. at $(\alpha_c,\beta_c)$ of  Eq.~(\ref{eq:cpp1p2051}). Furthermore,
by considering suitably short chains, see Appendix~\ref{sec:app1}, 
we were able to prove that there can be no factorizable states except if $p_1=p_2$.

The proof just referred to adds to the body of evidence displayed in the pertinent results of
Ref.~\onlinecite{hex15}, as well as in  Figs.~\ref{fig:sstp_abcl_st}
and~\ref{fig:sstp_acbcb_st}. 
All the above strongly suggest that for staggered chains the predictions of DW theory
regarding factorization, phase coexistence, and criticality come about only in an incipient 
and quantitatively approximate way.

Notwithstanding the statements just made, we note that  DW theory  for staggered chains gives reasonably good fits to the steady state exponential
 $\lambda_s$ in regions where it is  predicted to differ 
appreciably from zero, see data for $(\alpha,\beta)=(0.1,0.22)$ and $(0.3,0.115)$.
Additionally,  the DW theory predictions extracted from 
Eqs.~(\ref{eq:diff1})--(\ref{eq:diff2}), regarding $\ell$-independent difference between
sublattice steady state profiles, are well verified by numerics;
see Figs.~\ref{fig:sstp_acbc_st}--~\ref{fig:sstp_acbcb_st}. In Fig.~\ref{fig:sstp_acbc_st}
one can also see  evidence of the fast processes having almost completely died out at $t=240$,
though convergence toward  steady state takes significantly longer.

Concerning the DW description of the approach to steady state for staggered chains,
its approximate character is well illustrated by the data shown in Fig.~\ref{fig:nsp_abncl}.
There, we have been able to find a point on the phase diagram, rather close to but not {\em on} the
predicted coexistence line, where the vanishing of the gap is verified to rather good accuracy. 
Similarly, the numerically-evaluated coefficient of the leading finite-size correction is within
$8\%$ of the DW theory prediction. For an example of a case where the gap is definitely nonzero,
namely  $(\alpha,\beta)=(0.1,0.22)$ the numerical data exhibit rather broad error bars; nevertheless,
with the help of some plausible assumptions one can conclude (see Fig.~\ref{fig:nsp_abclp}) that the limiting gap value is of the same order of magnitude as predicted by theory, apart from a factor of order $2$ at most.

Before concluding, some further points merit discussion here. The main 
one concerns length and time scales. For the applicability of the macroscopic
view underlying the DW approach  one needs an appropriate separation and
ordering of such scales, particularly length scales (system size, domain
size, domain wall width, and lattice spacing). Such requirements are
typically well satisfied in the uniform chain systems investigated here. But
for the staggered chains the two-sublattice feature makes them questionable.
This may well be the basic reason for the imperfect DW account of this case.

The time scales are important not only for the theory but also in guiding
simulations. These scales are set by diffusion rates and bias velocities,
together with characteristic lattice or system lengths in gapped or gapless
spectra, and lattice traversal times (from real and imaginary parts of
the decay rate $R_1$). The latter ballistic effects are evident in certain of 
the current investigations (e.g. in filling of the lattice from the injection 
side in the
case of empty lattice initial conditions, see Fig.~\ref{fig:sstp_acbc_st} above), 
but have already been detailed  elsewhere (see, e.g., Figs.~2, 3, and~5
of Ref.~\onlinecite{sa02}, as well as Figs.~2 and~7
of  Ref.~\onlinecite{hex15}).

\begin{acknowledgments}
We thank Anatoly Kolomeisky for helpful discussions.
S.L.A.d.Q. thanks the Rudolf Peierls Centre for Theoretical Physics, Oxford, for
hospitality during his visit. The research of S.L.A.d.Q. is supported by 
the Brazilian agencies Conselho Nacional de Desenvolvimento Cient\'\i fico e
Tecnol\'ogico  (Grant No. 303891/2013-0) 
and Funda\c c\~ao de Amparo \`a Pesquisa do Estado do Rio
de Janeiro (Grants Nos. E-26/102.760/2012, E-26/110.734/2012, 
and E-26/102.348/2013).
\end{acknowledgments}
 
 \appendix

\section{Investigation of possibly factorized steady states}
\label{sec:app1}

We investigate the possibility of factorized steady states for the staggered chain TASEP using
direct application of the transition matrix. The method is feasible for open boundary systems
of small size.

We consider systems with $N$ sites, hence $2^N$ possible configurations, at small $N$.

We can write down the $2^N \times 2^N$ transition matrix $\cal W$ whose elements
${\cal W}_{ij}$ give the rates of transition from configuration $j$ to configuration $i$.
The ${\cal W}_{ij}$  are functions of the boundary injection and ejection rates $(\alpha,\beta)$
and of the internal hopping rates: $p$ for the uniform case, or $p_1$ and $p_2$ for 
the staggered case.

Any state can be written as a column vector in which the $i^{\rm th}$ element, $u_i$ say,
is the probability of configuration $i$. Steady states have vectors which are 
eigenvectors of $\cal W$ with zero eigenvalue. So the possibility of a steady state 
with any sort of factorization can be tested by applying $\cal W$ to its column vector.

For a fully factorizable state the $u_i$ can be written in the form $x^n\,y^{N-n}$ where $n$
is the number of particles in configuration $i$, and $x+y=1$.

It is easy to check that, for the uniform case at small $N$, such a state is indeed a steady state
subject to $\alpha$, $\beta$, $p$ satisfying $\alpha+\beta=p$ and to having
$x=1-y=\alpha/(\alpha+\beta)$.

Of course, the exact steady-state solution of the TASEP on a uniform chain~\cite{derr93}
already includes this result. However, nothing comparable is known for the staggered case.

In the latter case the DW approach suggests a steady state which is factorizable on each
sublattice, for $a+b=1$, where $a=\alpha/p_2$, $b=\beta/p_1$.

To verify or refute this, our procedure will be to apply the staggered-chain transition matrix to a
corresponding column vector having elements which are products of $x$, $y$, $X$, $Y$, with
$x$, $y$ corresponding to particle or vacancy at an odd-index site and $X$, $Y$ likewise
for even sites. 

Already one obtains conclusive results from size $N=3$. This involves an $8 \times 8$
dynamic matrix $\cal W$ which has off-diagonal elements $\alpha$, $\beta$, $p_1$, $p_2$ or zero,
and diagonal elements such that all column sums are zero.
Written on a basis in which the first vector element corresponds to all sites occupied, the second
element has the first two sites occupied and the last empty, and so on until the last element
corresponds to all sites empty, ${\cal W}$  is given by:

\begin{equation}
\begin{pmatrix}
{\!-\beta} & {0} & {0} &{\alpha} & {0} & {0} & {0} & {0} \cr
{\beta} & {\!-p_2\!} & {0} &{0} & {0} & {\alpha} & {0} & {0} \cr
{0} & {p_2} & {-p_1\!-\!\beta} &{ 0} & {0} & {0} & {\alpha} & {0} \cr
{0} & {0} & {p_1} &{-\alpha\!-\!\beta} & {0} & {0} & {0} & {0} \cr
{0} & {0} & {\beta} &{0} & {\!-p_1\!} & {0} & {0} & {\alpha} \cr
{0} & {0} & {0} &{\beta} & {p_1} & {-p_2\!-\!\alpha} & {0} & {0} \cr
{0} & {0} & {0} &{0} & {0} & {p_2} & {-\alpha\!-\!\beta} & {0} \cr
{0} & {0} & {0} &{0} & {0} & {0} & {\beta} & {\!-\alpha} 
\end{pmatrix}\ \  .
\label{eq:fake}
\end{equation}

The application of $\cal W$ to the state which is factorizable on each sublattice separately has to give a zero vector for that state to be a steady state. The resulting vector is
\begin{equation}
\begin{pmatrix}
{xX(\alpha y-\beta x)}\cr 
{X [\beta x^2-p_2xy+\alpha y^2 ]}\cr
{x[p_2yX+\alpha yY-(p_1+\beta)xY]}\cr
{x[p_1xY-(\alpha+\beta)yX]}\cr
{Y[\beta x^2-p_1 xy+\alpha y^2]}\cr
{y[\beta xX+p_1 xY-(p_2+\alpha)yX ]}\cr
{y[p_2 yX-(\alpha+\beta)xY]}\cr
{yY(\beta x-\alpha y)}
\end{pmatrix}\ \ .
 \label{eq:vector}
 \end{equation}
 So these elements are all zero only if
 \begin{equation}
 p_1=p_2 =\alpha+\beta\ ;\  x=X=\frac{\alpha}{\alpha+\beta}\ ;\  y=Y=\frac{\beta}{\alpha+\beta}\ .
 \label{eq:stag_cond}
 \end{equation}
 This means that the staggered chain with $p_1 \neq p_2$ has no factorizable states, not even with
 factorization on each sublattice separately.

\end{document}